\documentclass{article}
\usepackage[margin=1in]{geometry}
\usepackage{amsmath,amssymb}
\usepackage{enumerate}
\usepackage{graphicx}
\usepackage{xcolor}
\usepackage{natbib}

\renewcommand{\Pr}{\mbox{P}}

\usepackage[plain,noend]{algorithm2e}

\author{J. L. Wadsworth\\
\\
Department of Mathematics and Statistics, Fylde College, Lancaster University, \\LA1 4YF, U.K.}

\title{On the occurrence times of componentwise maxima and bias in likelihood inference for multivariate max-stable distributions}

\begin{document}

\maketitle

\begin{abstract}
Full likelihood-based inference for high-dimensional multivariate extreme value distributions, or max-stable processes, is feasible when incorporating occurrence times of the maxima; without this information, $d$-dimensional likelihood inference is usually precluded due to the large number of terms in the likelihood. However, some studies have noted bias when performing high-dimensional inference that incorporates such event information, particularly when dependence is weak. We elucidate this phenomenon, showing that for unbiased inference in moderate dimensions, dimension $d$ should be of a magnitude smaller than the square root of the number of vectors over which one takes the componentwise maximum. A bias reduction technique is suggested and illustrated on the extreme value logistic model.
\end{abstract}

\section{Introduction}
Let $X_i=(X_{i,1},\ldots,X_{i,d})$ $(i=1,\ldots,n)$, denote a collection of independent and identically distributed random vectors with standard Fr\'{e}chet margins, $\Pr(X_{i,j} \leq x_j) = \exp(-1/x_j)$, $x_j>0$ $(j=1,\ldots,d)$. Multivariate max-stable distributions arise as the only possible non-degenerate limits of suitably normalized componentwise maxima of such vectors. That is, if the normalized componentwise maximum vector,
\begin{align}
 M_n /n = \left(\max_{1\leq i \leq n} X_{i,1}, \ldots, \max_{1\leq i \leq n} X_{i,d}\right) / n, \label{eq:Mn}
\end{align}
satisfies
\begin{align}
  \Pr(M_n/n \leq x) \to G(x), ~~~x=(x_1,\ldots,x_d), \label{eq:dfconv}
\end{align}
as $n\to\infty$ for some non-degenerate $G$, then $G(x)=\exp\{-V(x)\}$, $x>0$, is a multivariate max-stable distribution, with standard Fr\'{e}chet margins. In~\eqref{eq:dfconv}, and throughout, inequalities between vectors should be interpreted componentwise.  The function $V$ describes the limiting dependence between maxima; it is homogeneous of order $-1$ and satisfies the marginal condition $V(\infty,\ldots,\infty,x_j,\infty,\ldots,\infty)=x_j^{-1}$ $(j=1,\ldots,d)$.

Full likelihood-based inference for multivariate max-stable distributions is inhibited by the fact that the $d$-dimensional density,
\begin{align}
 \frac{\partial^d}{\partial x_1 \cdots \partial x_d} e^{-V(x)},\label{eq:dens}
\end{align}
is a sum of the $d$th Bell number, $B_d$, of terms, each term corresponding to one possible partition of the set $\{1,\ldots,d\}$. When dimension $d$ is moderate to large, enumeration of all $B_d$ terms needed in the likelihood~\eqref{eq:dens} is infeasible. One solution is the use of pairwise composite likelihoods \citep{Padoan10}, but whilst this provides consistent inference, its efficiency can be low in comparison to estimators from full likelihoods \citep{Huser14}. However, \citet{StephensonTawn05} showed that if the occurrence times of the maxima were known, then the likelihood~\eqref{eq:dens} could be substantially simplified, since only one set of derivatives of $V$, corresponding to the partition detailing which maxima occurred in which original vectors, need be included. To make this concrete, suppose that $d=4$, and that the first two componentwise maxima occurred in a single vector, $X_a$, whilst the third and fourth occurred in two separate vectors, $X_b$, and $X_c$. Then $M_n = (X_{a,1}, X_{a,2}, X_{b,3}, X_{c,4})$, and the partition corresponding to these occurrence times is $\{\{1,2\},\{3\},\{4\}\}$. The limiting joint density of both the maxima and these occurrence times is
\[
 -V_{\{1,2\}}(x)V_{\{3\}}(x) V_{\{4\}}(x) e^{-V(x)},~~~ x=(x_1,x_2,x_3,x_4),
\]
where here and throughout subscripts of sets containing natural numbers denote partial differentiation with respect to the variables indexed by the subscript. \citet{WadsworthTawn14} used ideas of \citet{StephensonTawn05} to facilitate $d$-dimensional inference for a class of tractable max-stable processes, observing large efficiency gains. For relatively strong dependence their estimation seemed approximately unbiased, but potential bias was noted under weaker dependence. For a related class of models \citet{ThibaudOpitz14} also record lower bias from pairwise than full likelihoods in certain senarios, and  \citet{Huser14} note a similar effect for a high-dimensional extreme value logistic distribution. This note explores this effect, and in particular investigates how bias scales with dimension. A bias reduction technique, which keeps the number of terms in the likelihood manageable for moderate $d$, is suggested and illustrated on the extreme value logistic distribution.

\section{Independence}
\label{sec:Indp}
To fix ideas, consider the simplest max-stable dependence structure: independence. By \citet{StephensonTawn05}, if $X_i = (X_{i,1},\ldots,X_{i,d})$ $(i=1,\ldots, n)$ are random vectors that converge to the independence max-stable limit, then, with probability 1, the limiting partition of the occurrence times of the normalized componentwise maxima is $\{\{1\},\{2\},\ldots,\{d\}\}$, i.e., the relevant likelihood contribution is 
\[
 (-1)^d V_{\{1\}}(x)V_{\{2\}}(x) \cdots V_{\{d\}}(x) e^{-V(x)},~~~ x=(x_1,\ldots,x_d).
\]
A different partition of the maxima would indicate that the components were not mutually independent, and so bias the inference. Now consider the situation where $X_i = (X_{i,1},\ldots,X_{i,d})$, are exactly independent, both over $j=1,\ldots, d$ and $i=1,\ldots, n$. The data are already max-stable, so no dependence structure convergence need take place. However, the probability that in the componentwise maximum vector, each component $M_{n,j}$, $j=1,\ldots,d$, comes from a different underlying vector $X_i$ is
\[
 \frac{n}{n} \times \frac{n-1}{n} \times \frac{n-2}{n} \times \cdots \times  \frac{n-(d-1)}{n} = \frac{n!}{(n-d)! n^d}.
\]
This does indeed converge to 1 as $n\to \infty$, but unlike the dependence structure itself, which is already max-stable, some convergence must occur for the partition of the occurrence times to indicate independence. Moreover, if $d$ grows with $n$, then convergence of the partition of occurrence times to the correct one may be destroyed. One typically thinks of $d$ as fixed, but in a spatial application where data are limited, and the number of sites is relatively high, $d$ may well be comparable to a power of $n$. Using Stirling's formula, as $n\to \infty$,
\begin{align}
 \frac{n!}{(n-d)! n^d} &=  e^{-d} \left(1-\frac{d}{n}\right)^{-1/2-(n-d)} + O(n^{-1}). \label{eq:dn}
\end{align}
If $d=d_n \to \infty$ is an integer sequence changing with $n$ such that $d_n=o(n)$ then~\eqref{eq:dn} remains valid, and can be expressed as
\begin{align}
    \exp\left[-\frac{d_n^2}{2n}\left\{1+o(1)\right\}\right], ~~n\to\infty,\label{eq:dn2}
\end{align}
so that the condition for the limit to equal 1 is $d_n=o(n^{1/2})$.

\section{General max-stable dependence}
\label{sec:General}
We remain in the max-stable framework, but consider more general dependence structures. Now suppose that $X_i = (X_{i,1},\ldots,X_{i,d})$ $(i=1,\ldots, n)$ come from a multivariate max-stable distribution with standard Fr\'{e}chet margins, so that $\Pr(X_i\leq x) = F(x) = e^{-V(x)}$, $x=(x_1,\ldots,x_d)>0$. We assume that the joint density exists, and can be expressed as
\begin{align}
 e^{-V(x)}  \sum_{k=1}^d \sum_{j=1}^{\{d,k\}} \prod_{l=1}^k \{-V_{\pi_{k,j,l}}(x)\}, \label{eq:Vdens}
\end{align}
where the $\{d,k\}$ are Stirling numbers of the second kind, i.e., the number of the $B_d$ partitions composed of $k$ sets, and $\{\pi_{k,j,l}\}_{l=1}^k$ is an enumeration of all subsets in the $j$th partition of size $k$. Max-stability of the $X_i$ implies that $M_n / n$ is equal in distribution to $X$, as
\[
 \Pr(M_n/n \leq x) = F(nx)^n = e^{-nV(nx)} = F(x).
\]
The joint density function, $f_X$, can thus be expressed in two equivalent ways, as
\begin{align}
  \frac{\partial^d}{\partial x_1 \cdots \partial x_d} F(x) =  n^d \frac{\partial^d}{\partial nx_1 \cdots \partial nx_d} F(nx)^n.~\label{eq:diff}
\end{align}
Evaluation of these derivatives requires repeated use of the product rule, but the natural decomposition for the functions in the product is different. It turns out that the two resulting summations highlight the partitions of occurrence of componentwise maxima from an infinite and finite number of samples, respectively.

Let $R_n$ be a sequence of random variables taking values on the space of partitions of $\{1, \ldots, d\}$, which represent the events that the componentwise maxima from $n$ independent vectors come from a particular configuration of the $n$ vectors. Denote by $R$ the limiting random variable, i.e., $R_n$ converges in distribution to $R$ as $n\to\infty$, and let $f_{X,R_n}$ and $f_{X,R}$ denote the joint densities of $(X,R_n)$ and $(X,R)$ respectively. The left-hand side of~\eqref{eq:diff} is most simply expressed by~\eqref{eq:Vdens}, whilst the right-hand side has the natural expression
\begin{align}
 f_X(x) &= n^d \sum_{k=1}^d \sum_{j=1}^{\{d,k\}} \frac{n!}{(n-k)!} F(n x)^{n-k} \prod_{l=1}^k F_{\pi_{k,j,l}}(n x) \label{eq:STdecomp}\\
 & = n^d \sum_{k=1}^d \sum_{j=1}^{\{d,k\}} \frac{n!}{(n-k)!} e^{-(n-k)V(nx)} \prod_{l=1}^k \left\{e^{-V(nx)}\right\}_{\pi_{k,j,l}} \label{eq:STV}\\ 
 &= \sum_{k=1}^d \sum_{j=1}^{\{d,k\}} f_{X,R_n}(x; \{\pi_{k,j,l}\}_{l=1}^k), \label{eq:part}
\end{align}
which is formula~(2.4) of \citet{StephensonTawn05}. They argue for the equality between lines~\eqref{eq:STdecomp} and~\eqref{eq:part} as follows: when the partition is of size $k$, there are $n!/(n-k)!$ ways to select the $k$ vectors in which the maxima occur; $F(nx)^{n-k}$ is the probability that the remaining $n-k$ of the vectors take values less than the observed scaled maxima $nx$, whilst for the partition $\{\pi_{k,j,l}\}_{l=1}^k$, $F_{\pi_{k,j,l}}(nx)$ ($l=1,\ldots,k$) are the joint densities of the maxima and censored components, which are multiplied due to independence over the $n$ repetitions. Taking any particular partition, for fixed $d$ and as $n\to\infty$, 
\[
 f_{X,R_n}(x; \{\pi_{k,j,l}\}_{l=1}^k) \to f_{X,R}(x; \{\pi_{k,j,l}\}_{l=1}^k) =  e^{-V(x)} \prod_{l=1}^k \{-V_{\pi_{k,j,l}}(x)\}.
\]
It follows that~\eqref{eq:Vdens} may also be written 
\begin{align}
\sum_{k=1}^d \sum_{j=1}^{\{d,k\}} f_{X,R}(x; \{\pi_{k,j,l}\}_{l=1}^k). \label{eq:partlimit}                                      
\end{align}

Integration of $f_{X,R_n}$ or $f_{X,R}$ with respect to $x$ over $\mathbb{R}^d_+$ yields $\Pr(R_n= \{\pi_{k,j,l}\}_{l=1}^k)$ or $\Pr(R=\{\pi_{k,j,l}\}_{l=1}^k)$.  As in Section~\ref{sec:Indp}, consider the probability that all the maxima occur in separate events, i.e., $\Pr(R_n=\{\{1\},\ldots,\{d\}\})$. The contribution to summation~\eqref{eq:part} is
\begin{align*}
f_{X,R_n}(x; \{\{1\},\ldots,\{d\}\})&= \frac{n!}{(n-d)! n^d} \left\{(-1)^{d} V_{\{1\}}(x)V_{\{2\}}(x)\cdots V_{\{d\}}(x)\right\} e^{-V(x)},
\end{align*} 
so that
\begin{align}
 \Pr(R_n=\{\{1\},\ldots,\{d\}\}) = \frac{n!}{(n-d)! n^d}  \Pr(R=\{\{1\},\ldots,\{d\}\}). \label{eq:pr1}
\end{align}
Expression~\eqref{eq:pr1} shows that potential bias manifests itself in the same manner as in the independence case, where $\Pr(R=\{\{1\},\ldots,\{d\}\}) = 1$. Under asymptotic dependence, the particular event $R=\{\{1\},\ldots,\{d\}\}$ may in general have quite a low probability, even for weakly dependent processes, but similar issues arise for all partitions of size $m$, where $m$ is much closer to $d$ than to 1. By matching partitions in~\eqref{eq:part} and~\eqref{eq:partlimit} and isolating the relevant components of~\eqref{eq:Vdens} and~\eqref{eq:STV} one can establish that for $m\leq d-1$
\begin{align}
 \Pr(R_n= \{\pi_{m,j,l}\}_{l=1}^m) = \frac{n!}{(n-m)! n^m} \left\{ \Pr(R= \{\pi_{m,j,l}\}_{l=1}^m) + \sum_{k=m+1}^{d} \sum_{r=1}^{N_{j,k}} \frac{1}{n^{k-m}} \Pr(R= \{\pi_{k,r,l}\}_{l=1}^k)\right\},\label{eq:Rn}
\end{align}
with $N_{j,k}$ being the number of partitions of size $k$ induced by the partial differentiation operation $\prod_{l=1}^m \left\{e^{-V(x)}\right\}_{\pi_{m,j,l}}$. Bias can therefore be anticipated when $d$ is large for a given $n$, and the probabilities $\Pr(R= \{\pi_{m,j,l}\}_{l=1}^m)$, for $m$ near to $d$, are large compared to those for $m$ near to 1. Indeed, for $m=1$,~\eqref{eq:Rn} is
\begin{align*}
 \Pr(R_n= \{1,\ldots,d\}) =  \Pr(R= \{1,\ldots,d\}) + \sum_{k=2}^{d} \sum_{j=1}^{\{d,k\}} \frac{1}{n^{k-1}} \Pr(R= \{\pi_{k,j,l}\}_{l=1}^k), 
\end{align*}
so that the difference $\Pr(R_n= \{1,\ldots,d\})-\Pr(R= \{1,\ldots,d\})$ is bounded above by $1/n$. For strongly dependent processes it is these probabilities for $m$ near 1 that will dominate, thus bias will be much smaller.

\section{Second order bias reduction}
\label{sec:SecondOrder}
The calculations of Section~\ref{sec:General} suggest the use of the contributions $f_{X,R_n}$, in place of $f_{X,R}$, in a likelihood. However, from~\eqref{eq:STV} one can see that the contribution $f_{X,R_n}$ may once more be a sum of a large number of terms. A second order correction, i.e., including all terms of first and second order in $n$ from $f_{X,R_n}$ appears to be a viable alternative for moderate $d$. This is tantamount to assuming that realizations of the random variable $R_n$, whose probability mass function is described by equations such as~\eqref{eq:Rn}, are in fact realizations of the random variable $R_n^*$, which has probability mass function
\begin{align}
 \Pr(R_n^*= \{\pi_{m,j,l}\}_{l=1}^m) =  \Pr(R= \{\pi_{m,j,l}\}_{l=1}^m)\left\{1-\frac{m(m-1)}{2n}\right\} +  \sum_{r=1}^{N_{j,{m+1}}} \frac{1}{n} \Pr(R= \{\pi_{m+1,r,l}\}_{l=1}^{m+1}), \label{eq:Rnstar}
\end{align}
for $j=1,\ldots,\{d,m\}$, and $m=1,\ldots,d-1$; for $m=d$ the final summation is empty and is dropped. Equation~\eqref{eq:Rnstar} is simply the right hand side of~\eqref{eq:Rn} truncated after terms of order $1/n$. By contrast, when using the Stephenson--Tawn likelihood one assumes that the random variables $R_n$ are realizations of the limiting random variable $R$. We denote this second order truncated density by $f_{X,R_n^*}$. This is a valid density; because~\eqref{eq:part} is equal to~\eqref{eq:Vdens}, all terms of the same order in $n$ must cancel. However, there is a restriction on the dimension $d$ for use of $R_n^*$, since for positivity, one can see from~\eqref{eq:Rnstar} that we require $1-d(d-1)/(2n) >0$, i.e., $n>d(d-1)/2$. 

In general, if the size of the partition $|R_n|=m$, and if each of these $m$ subsets contains $d_i$ elements, $i=1,\ldots,m$, so that $\sum_{i=1}^m d_i = d$, then there will be $\sum_{i=1}^m \{d_i,2\}$ additional terms in the density contribution $f_{X,R_n^*}$ compared with $f_{X,R}$. Since $\{d,2\}= 2^{d-1}-1$, the additional number of terms is $\sum_{i=1}^m (2^{d_i-1}-1) \leq 2^{d-1}-1$. 

As a concrete example, suppose that $d=5$ and $R_n=\{\{1,2\},\{3,4\},\{5\}\}$, so that the leading order term corresponds to $-V_{\{1,2\}}V_{\{3,4\}}V_{\{5\}}$. The size of the partition is $m=3$, with $d_1=d_2=2$ and $d_3=1$. Then there are $(2^1-1) + (2^1-1) + (2^0-1) = 2$ terms of order $1/n$, which correspond to $V_{\{1,2\}}V_{\{3\}}V_{\{4\}}V_{\{5\}}$ and $V_{\{1\}}V_{\{2\}}V_{\{3,4\}}V_{\{5\}}$. Therefore in this case 
\begin{align*}
f_{X,R_n^*} &= \left[-V_{\{1,2\}}V_{\{3,4\}}V_{\{5\}} (1-3/n) +V_{\{1,2\}}V_{\{3\}}V_{\{4\}}V_{\{5\}}/n +V_{\{1\}}V_{\{2\}}V_{\{3,4\}}V_{\{5\}}/n\right]e^{-V}.
\end{align*}

We implement this idea for the logistic model \citep{Tawn90}, where $V(x) = (\sum_{i=1}^d x_i^{-1/\alpha})^{\alpha}$, $\alpha\in(0,1]$, using the likelihood contribution $f_{X,R_n^*}$ in place of $f_{X,R}$. Table~\ref{tab:bias} details the sample bias and standard deviation of 1500 maximum likelihood estimates of $\alpha$, each calculated from a sample of 100 $d$-vectors $M_n/n$, see equation~\eqref{eq:Mn}, with each $X_i$ simulated directly from the max-stable logistic model. The sizes of $n$, $d$, and level of dependence $\alpha$ varied as detailed in the table; values of $\alpha$ closer to 0 indicate stronger dependence. As expected, the results suggest that the largest reduction in bias occurs under weak dependence and for smaller $n$; as $n$ increases bias should disappear and both methods become equivalent. Standard deviations are smaller for the Stephenson--Tawn likelihood, though the difference is generally quite small.

The first three lines of Table~\ref{tab:nterms} display the mean number of terms used in the second order likelihood, for the ranges of $d$ and $\alpha$ used in Table~\ref{tab:bias}; the second three lines are discussed in Section~\ref{sec:Comments}. The figures are averaged across $n$, as there is little variation over $n$. For weaker dependence the numbers are much smaller, demonstrating that this bias reduction procedure is most feasible when it is most beneficial, at least in moderate dimensions.

\begin{table}[ht]
\small
\centering 
\caption{Top: Sample bias (top three rows) and sample standard deviation (second three rows) from the second order likelihood $f_{X,R_n^*}$; figures have been multiplied by 10000. An asterisk indicates that the bias is significantly different from zero at a 5\% level. Bottom: as top, but for the Stephenson--Tawn likelihood.}

\begin{tabular}{lrrrrrrrrrrrr}

 &\multicolumn{3}{c}{$\alpha=0.1$} &\multicolumn{3}{c}{$\alpha=0.4$}&\multicolumn{3}{c}{$\alpha=0.7$}&\multicolumn{3}{c}{$\alpha=0.9$}\\
   $n$ $|$ $d$  & 6 & 8 & 10 & 6 & 8 & 10 & 6 & 8 & 10 & 6 & 8 & 10 \\ 
 &  \multicolumn{12}{c}{Second order likelihood}\\

$50 $ & $ 1 $ & $ 0 $ & $ 0 $ & $ 1 $ & $ -2 $ & $ -1 $ & $ -5 $ & $ 10^* $ & $ 10^* $ & $ -5 $ & $ -25^* $ & $ -60^* $\\ 
$  100 $ & $ 0 $ & $ -1 $ & $ 0 $ & $ 2 $ & $ -3 $ & $ 2 $ & $ 1 $ & $ 5 $ & $ 2 $ & $ 2 $ & $ 5 $ & $ -6^* $\\ 
$  500 $ & $ 0 $ & $ 0 $ & $ -1 $ & $ 0 $ & $ 2 $ & $ -3 $ & $ -6 $ & $ -11^* $ & $ -3 $ & $ -3 $ & $ -1 $ & $ 1$ \\ 
$50 $ & $ 37 $ & $ 30 $ & $ 26 $ & $ 126 $ & $ 107 $ & $ 93 $ & $ 164 $ & $ 134 $ & $ 120 $ & $ 134 $ & $ 114 $ & $ 97$ \\ 
$  100 $ & $ 35 $ & $ 29 $ & $ 26 $ & $ 124 $ & $ 107 $ & $ 94 $ & $ 161 $ & $ 135 $ & $ 119 $ & $ 132 $ & $ 108 $ & $ 97$ \\ 
$  500 $ & $ 35 $ & $ 30 $ & $ 27 $ & $ 122 $ & $ 102 $ & $ 90 $ & $ 160 $ & $ 130 $ & $ 117 $ & $ 125 $ & $ 102 $ & $ 89$ \\ 

 & \multicolumn{12}{c}{Stephenson--Tawn likelihood}\\

$50 $ & $ 0 $ & $ -1 $ & $ -1$ & $ -30^* $ & $ -34^* $ & $ -34^* $ & $ -167^* $ & $ -178^* $ & $ -199^* $ & $ -351^* $ & $ -425^* $ & $ -490^*$ \\ 
$100 $ & $ -1 $ & $ -1 $ & $ 0 $ & $ -14^* $ & $ -19^* $ & $ -14^* $ & $ -82^* $ & $ -91^* $ & $ -106^* $ & $ -181^* $ & $ -220^* $ & $ -263^*$ \\ 
$500 $ & $ 0 $ & $ 0 $ & $ -1 $ & $ -3 $ & $ -1 $ & $ -6^* $ & $ -23^* $ & $ -30^* $ & $ -25^* $ & $ -41^* $ & $ -48^* $ & $ -56^*$ \\ 
  
$50 $ & $ 36 $ & $ 30 $ & $ 26 $ & $ 123 $ & $ 104 $ & $ 90 $ & $ 154 $ & $ 124 $ & $ 110 $ & $ 125 $ & $ 105 $ & $ 91$ \\ 
$  100 $ & $ 35 $ & $ 29 $ & $ 26 $ & $ 122 $ & $ 105 $ & $ 92 $ & $ 155 $ & $ 129 $ & $ 113 $ & $ 124 $ & $ 102 $ & $ 90$ \\ 
$  500 $ & $ 35 $ & $ 30 $ & $ 27 $ & $ 122 $ & $ 101 $ & $ 90 $ & $ 158 $ & $ 129 $ & $ 115 $ & $ 123 $ & $ 100 $ & $ 87$ \\ 
\end{tabular}
\label{tab:bias}
\end{table}
\normalsize

\begin{table}[ht]
\centering
\caption{Mean number of terms in the second order likelihood, averaged over $n=50,100,500$ for $d=6,8,10$; $n=200,500,1500$ for $d=15,20$, and $n=1500$ for $d=50$. Figures under 1000 are given to the nearest integer; figures over 1000 are given to three significant figures.}

\begin{tabular}{rrrrr}
$d$ & $\alpha=0.1$ & $\alpha=0.4$ & $\alpha=0.7$ & $\alpha=0.9$ \\ 
6 & 28 & 17 & 8 & 3 \\ 
  8 & 107 & 57 & 23 & 8 \\ 
  10 & 417 & 203 & 73 & 20 \\ 
 15 & $1.30\times 10^4$ & $5.32\times 10^3$ & $1.58\times 10^3$ & $354$ \\ 
  20 & $3.98\times 10^5$  & $1.48\times 10^5$  & $3.91\times 10^4$  & $8.02\times 10^3$  \\ 
  50 & $3.84\times 10^{14}$  & $1.06\times 10^{14}$  & $1.99\times 10^{13}$  & $3.54\times 10^{12}$  \\
\end{tabular}
\label{tab:nterms}
\end{table}

 In order to assess the relative effects of convergence of the dependence structure and of the partition variable, we also consider inference on data in the domain of attraction of the logistic max-stable distribution. Specifically, we use data with standard Fr\'{e}chet margins and Archimedean dependence structure, so that
 \[
  F(x) = \phi\left[\sum_{i=1}^d\phi^{-1}\{\exp(-1/x_i)\}\right],
\]
 where the Archimedean generator is $\phi(x) = (x^{\alpha}+1)^{-1}$, $\alpha\in(0,1]$, termed an outer power Clayton copula, and simulated using the copula library in R \citep{HofertMaechler11,Hofert14}. These data are in the domain of attraction of the logistic distribution with the same parameter value $\alpha$. Table~\ref{tab:Arch} details the sample bias from 1500 maximum likelihood estimates each calculated from 100 $d$-vectors $M_n/n$. Standard deviations give a broadly similar picture to Table~\ref{tab:bias}, and hence are omitted. As should be expected the biases are all more severe than in Table~\ref{tab:bias}, since there is also dependence structure convergence involved, but the bias is reduced by use of the second order likelihood, as it alleviates one of the causes. The mean numbers of terms in the likelihood do not differ by more than two from the corresponding figures in Table~\ref{tab:nterms}.


\begin{table}[ht]
\centering
\caption{Biases as in Table~\ref{tab:bias}, but for data simulated from the outer power Clayton copula.}
\begin{tabular}{rrrrrrrrr}
 &\multicolumn{2}{c}{$\alpha=0.1$} &\multicolumn{2}{c}{$\alpha=0.4$}&\multicolumn{2}{c}{$\alpha=0.7$}&\multicolumn{2}{c}{$\alpha=0.9$}\\
$n$ $|$ $d$ & 6 & 10 & 6 & 10 & 6 & 10 & 6 & 10 \\ 
    & \multicolumn{8}{c}{Second order likelihood}\\
50    & $ -11^* $ & $ -11^* $ & $ -61^* $ & $ -58^* $ & $ -158^* $ & $ -194^* $ & $ -277^* $ & $ -422^*$ \\ 
  100  & $ -5^* $ & $ -5^* $ & $ -30^* $ & $ -32^* $ & $ -85^* $ & $ -106^* $ & $ -161^* $ & $ -214^*$ \\ 
  500  & $ -3^* $ & $ 0 $ & $ -4 $ & $ -6^* $ & $ -21^* $ & $ -26^* $ & $ -37^* $ & $ -52^*$ \\  

    &  \multicolumn{8}{c}{Stephenson--Tawn likelihood}\\
  50  & $ -12^* $ & $ -11^* $ & $ -90^* $ & $ -88^* $ & $ -315^* $ & $ -385^* $ & $ -634^* $ & $ -835^*$ \\ 
  100  & $ -5^* $ & $ -5^* $ & $ -45^* $ & $ -48^* $ & $ -167^* $ & $ -211^* $ & $ -353^* $ & $ -477^*$ \\
  500  & $ -3^* $ & $ 0 $ & $ -7^* $ & $ -9^* $ & $ -38^* $ & $ -48^* $ & $ -76^* $ & $ -111^*$ \\ 

\end{tabular}
\label{tab:Arch}
\end{table}

\section{Comments}
\label{sec:Comments}
The inclusion of information on occurrence times, which makes likelihood inference for max-stable distributions feasible in high dimensions, can also lead to estimation bias if the dimension is too high for the number of events over which one maximizes. In practice there could be situations where we cannot maximize over 100 events, perhaps one summer's worth of daily maximum temperatures, and perform 10-dimensional inference using the Stephenson--Tawn likelihood, expecting it to be unbiased. Bias is not the only concern in the performance of an estimator, but efficiency considerations have been explored more thoroughly in \citet{WadsworthTawn14}, \citet{ThibaudOpitz14} and~\citet{Huser14}. The latter empirically decompose root mean squared errors of several likelihood estimators into bias and variance contributions, thus offering some guide as to when bias may dominate.

The simulation study of Section~\ref{sec:SecondOrder} was restricted to $d\leq 10$; in spatial applications, higher dimensions frequently arise. When $d$ is much larger than $10$, the number of terms in the second order likelihood can become prohibitive even under weak dependence. The final three lines of Table~\ref{tab:nterms} detail such numbers, which were calculated, but not enumerated, via simulation as per the first three lines. As noted in Section~\ref{sec:SecondOrder}, the second order correction requires $n>d(d-1)/2$, and thus in order to use $f_{X,R_n^*}$, $n$ need be of order $d^2$, which is also increasingly unrealistic as $d$ grows. Interestingly however, as $d$ and $n$ grow simultaneously in the manner $d \asymp n^{1/2}$, the overall bias from the Stephenson--Tawn likelihood for the logistic model decreases. For $\alpha=0.9$, $d=10,20,30,40$ and $n=d^2/2$, the biases, multiplied by 10000, from 1500 repetitions are respectively $-487, -220, -136, -96$, with similar behaviour for other values of $\alpha$.  This pattern of reduction in bias also extends to $n$ growing more slowly with $d$; for $n=2d$, the equivalent figures are $-1030, -877, -798, -744$ . Tables~\ref{tab:bias} and~\ref{tab:Arch}, and equations~\eqref{eq:dn2} and~\eqref{eq:Rn} clearly indicate that for a fixed $n$, bias will increase in $d$ under weak dependence. Whilst equations~\eqref{eq:pr1} and~\eqref{eq:Rn} indicate that for $d=\Omega(n^{1/2})$ the ratio $\Pr(|R_n|=m)/\Pr(|R|=m)$ does not converge to unity for $m$ near $d$, the behaviour of $\Pr(|R_n| = m)$, and its effect on the maximum likelihood estimate, becomes difficult to analyze fully as the dimension grows. Further work could explore this, however the fact remains that in practice one may seldom have $n$ sufficiently large to rely on bias disappearing.

Estimation of model parameters is often not the primary goal of an analysis, but rather estimation of probabilities of extreme events. In a multivariate setting, there are several possibilities for defining extreme events, and the context of application will normally dictate those of interest. As one example, let $x_p$ be the level exceeded by at least one component of $M_n/n$ with probability $p$. For the logistic model the maximum likelihood estimate of $x_p$ is $\hat{x}_p = d^{\hat{\alpha}}/\{-\log(1-p)\}$, which is negatively biased if the maximum likelihood estimate $\hat{\alpha} < \alpha$, and has true non-exceedance probability of $(1-p)^{d^{\alpha-\hat{\alpha}}}$.

We focussed on the likelihood for the componentwise maximum distribution. Provided that all observations are available then a threshold-based model is usually preferred. Similar issues to those discussed herein apply, though the partition variables in question will be of a different nature and indicate locations of threshold exceedances rather than configurations of maxima. In fact the \citet{StephensonTawn05} likelihood for maxima is a special case of a threshold-based likelihood, with the threshold set at the observed maxima \citep{WadsworthTawn14}.

The discussion above suggests possibilities for bias reduction techniques. Implementing one such possibility for one simple model indicated that indeed bias reduction can be achieved, with the largest gains for weak dependence. Whilst the second order likelihood becomes difficult to implement in higher dimensions, the problem of bias appears to reduce for $d$ and $n$ growing simulataneously. A more thorough investigation into different likelihoods and for more realistic models seems warranted.

\subsection*{Acknowledgement}

I thank Jonathan Tawn and Rapha\"{e}l Huser for helpful comments and discussion, and two referees for constructive suggestions that have improved the article. This work was done whilst based at the Statistical Laboratory, University of Cambridge, and Lancaster University.

\bibliographystyle{apalike}
\bibliography{PartitionBib}
\end{document}